\documentclass[11pt,twoside]{article}

\usepackage{asp2006}
\usepackage{fancyhdr,graphicx,caption,rotating,multirow,lscape}
\DeclareGraphicsExtensions{.eps,.eps.gz,.ps,.jpg,.tiff}

\begin{document}

\title{On the influence of blends on the potential of ground-based transit surveys}  


\author{M. Gillon}
\affil{Observatoire de Gen\`eve, 51 Chemin des Maillettes, 1290 Sauverny,
Switzerland }
\affil{Institut d'Astrophysique et de G\'eophysique,  Universit\'e
  de Li\`ege,  All\'ee du 6 Ao\^ut, 17,  Bat.  B5C, Li\`ege 1, Belgium}
\author{P. Magain}
\affil{Institut d'Astrophysique et de G\'eophysique,  Universit\'e
  de Li\`ege,  All\'ee du 6 Ao\^ut, 17,  Bat.  B5C, Li\`ege 1, Belgium}



\begin{abstract} 
Transit surveys have to observe many stars all at once in order to compensate for the rarity of the searched events. Such surveys, especially the ones observing a deep field of view and/or broadening their stellar images, have to deal with a relatively high level of crowding. This crowding could lead to a reduction of the number of detectable transits, and the estimation of the potential of such surveys without taking into account the influence of blends could give overoptimistic results. We have developed a code which allows to estimate the extent by which such a survey is affected by the crowding of the field of view. Our results show that the influence of blends is important only for severe levels of crowding and is in general much less crucial than the influence of red noise. 
\end{abstract}

\section{Introduction}

As shown by Pont et al. (this volume), a partial explanation for the disagreement between former expectations and the actual harvest of transit surveys is that most of the computations did not take into account the influence of systematics (red noise). A complementary explanation for this disagreement could come from the fact that the crowding of the field of view is generally not taken into account in these computations. Nevertheless, transit surveys are searching for rare events and have thus to observe thousands of stars in the same field of view. Deep surveys as OGLE-III (Udalski et al. 2002) achieve this requirement by observing crowded fields in the Galactic plane or towards the Bulge, while shallow surveys as HAT (Bakos et al. 2004) observe less dense fields but have generally to widen their PSF to avoid saturation and to keep a relatively good spatial sampling in spite of their large pixel size. The influence of blends should thus  not be \emph{a priori} neglected in any potential estimation aiming at realistic results. 

In a previous analysis (Gillon et al. 2005, hereafter G05), we have presented a method designed to carry out a comparative analysis of transit surveys by computing their expected harvest under given assumptions. We have improved  this method by including the influence of the red noise and the necessity to confirm a detection with the radial velocity method, and also by modelling the influence of the crowding on the detection potential.  We present here some results obtained with this method for several ground-based surveys in an attempt to compare the influence of the blends with the one of systematics (Sect. 3). The modelling of the influence of blends is briefly presented in Sect. 2. Section 4 summarizes the main results. 
   
\section{The method}

The method used in this analysis is an extension of the  one developed in G05, i.e. it consists in the computation of the final harvest of a given survey, taking into account its observational and instrumental parameters and using simple work assumptions. This new method is described in details in Gillon et al. (2007, hereafter G07). Here we describe briefly the way the influence of the crowding of the field and of the red noise are taken into account. 

\subsection{Influence of  the crowding}

The simple fact that the PSF of a given star in the focal plane has a spatial extension leads, if other stars are present in the field of view, to the possibility that its PSF is contaminated by the PSFs of other stars.
 If we assume that a blended star undergoes a planetary transit during the observation, we will have a decrease of the S/N due to the combination of 2 effects: the  \emph{signal dilution} and the \emph{noise increase}.

The \emph{signal dilution} comes from the fact that the signal itself consists of a relative dimming of the brightness of an object. If the object is an isolated star, the amplitude of the signal is given by (neglecting limb-darkening):
\begin{equation}
\frac{\Delta L_{\star}}{L_{\star}} = \bigg(\frac{R_p}{R_{\star}}\bigg)^2\textrm{ ,}
\end{equation} where $L_{\star}$ is the luminosity of the star, $R_{\star}$ its radius and $R_p$ the radius of the planet. If a fraction of the luminosity of neighbouring stars $L_{blend}$ is added to the luminosity  $L_{\star}$, the amplitude of the signal becomes: 
 \begin{equation}
\frac{\Delta L_{\star}}{(L_{\star}+L_{blend})} < \bigg(\frac{R_p}{R_{\star}}\bigg)^2\textrm{ .}
\end{equation}  The signal becomes fainter and the planet is more difficult to detect. Nevertheless, this effect has an influence only if  simple aperture photometry is used. When a more sophisticated reduction method is used, as a PSF fitting algorithm, this effect generally plays no role. An exception is the case of 2 or more stars which cannot be separated by the detection algorithm. 

The \emph{noise increase} is due to the fact that the light coming from blending stars has its own photon noise, which adds to the noise of the target star. Furthermore, the blending stars have their own level of variability which will add another source of noise. In the case of ground-based observations,  they also bring a noise contribution due to the scintillation.

Aside from these two harmful effects, we have to take into account the fact that the crowding also influences the \emph{detection} of target stars by the photometric algorithm (see G07 for details).

Our algorithm takes into account every particular case of blend by giving it a statistical weigth and by computing the corresponding minimal planetary radius for detection. The latter is directly obtained by the computation of the S/N for each case. 

\subsection{Influence of the red noise}

As shown clearly by Pont, Zucker \& Queloz (2006, hereafter PZQ), all the existing transit surveys suffer from the presence of residual systematics in their photometry, even after detrending with sophisticated algorithms. For each survey considered, we repeat our computations for 4 different levels of red noise (represented by the parameter $\sigma_r$ = 0, 1, 2 and 3 mmag) using the simplest of the formulations presented in PZQ.

\section{Results}

\subsection{Wide shallow survey}

We compute the harvest of a fictitious wide shallow survey considered as representative of the many existing surveys of this kind, observing in the $I$ filter a 60 deg$^2$ field of view  in the direction of the Kepler field ($b=76.3^\circ $, $l=13.5^\circ $) during 1 month with a 10 cm telescope. We assume that aperture photometry is used, with an aperture computed to contain 95 \% of the stellar flux, and a broadening of the PSF width to 28 arcsec. We use a window function  taking into account the meteorological conditions and assume a  favorable location. We also compute the harvest of the same survey, but assuming a continuous window function, as could be obtained from Dome C (Fressin et al. 2005). Our results are summarized in Figs. 1 and 2, which show that: \begin{itemize}
\item Our fictitious survey finds more easily planets transiting many times during the run, i.e. Very Hot Jupiters ($VHJ$), even if they are much rarer than Hot Jupiters ($HJ$, assumed to be 5 times more frequent than  $VHJ$). 
\item The fraction of planets not detected because of blends is lower than 10 \%, while the influence of red noise is much stronger, especially when its amplitude is larger than 1 mmag. 
\item A continuous window function has a large impact on the final harvest. In this case, the harvest is more then 10 times larger from DomeC.
\end{itemize}

\begin{figure}[ht!]
\centering
\includegraphics[angle=0,width=9cm]{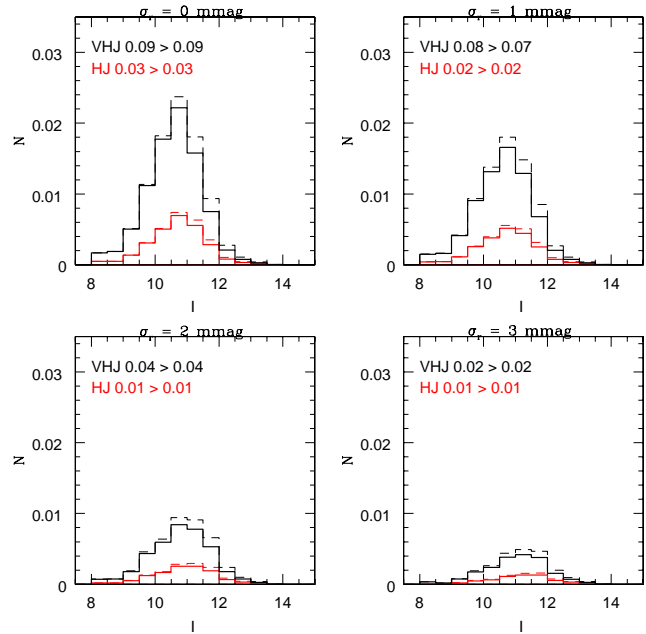}
\caption{Distribution of the planets detected by a fictitious wide shallow survey observing from a favorable location as a function of the $I$ magnitude, for 4 different levels of red noise. $Dotted$ $lines$: influence of blends neglected, $solid$ $lines$: influence of blends taken into account. $HJ$: Hot Jupiters (period between 3 and 9 days), $VHJ$: Very Hot Jupiters (period between 1 and 3 days). }
\end{figure}

\begin{figure}
\centering
\includegraphics[angle=0,width=9cm]{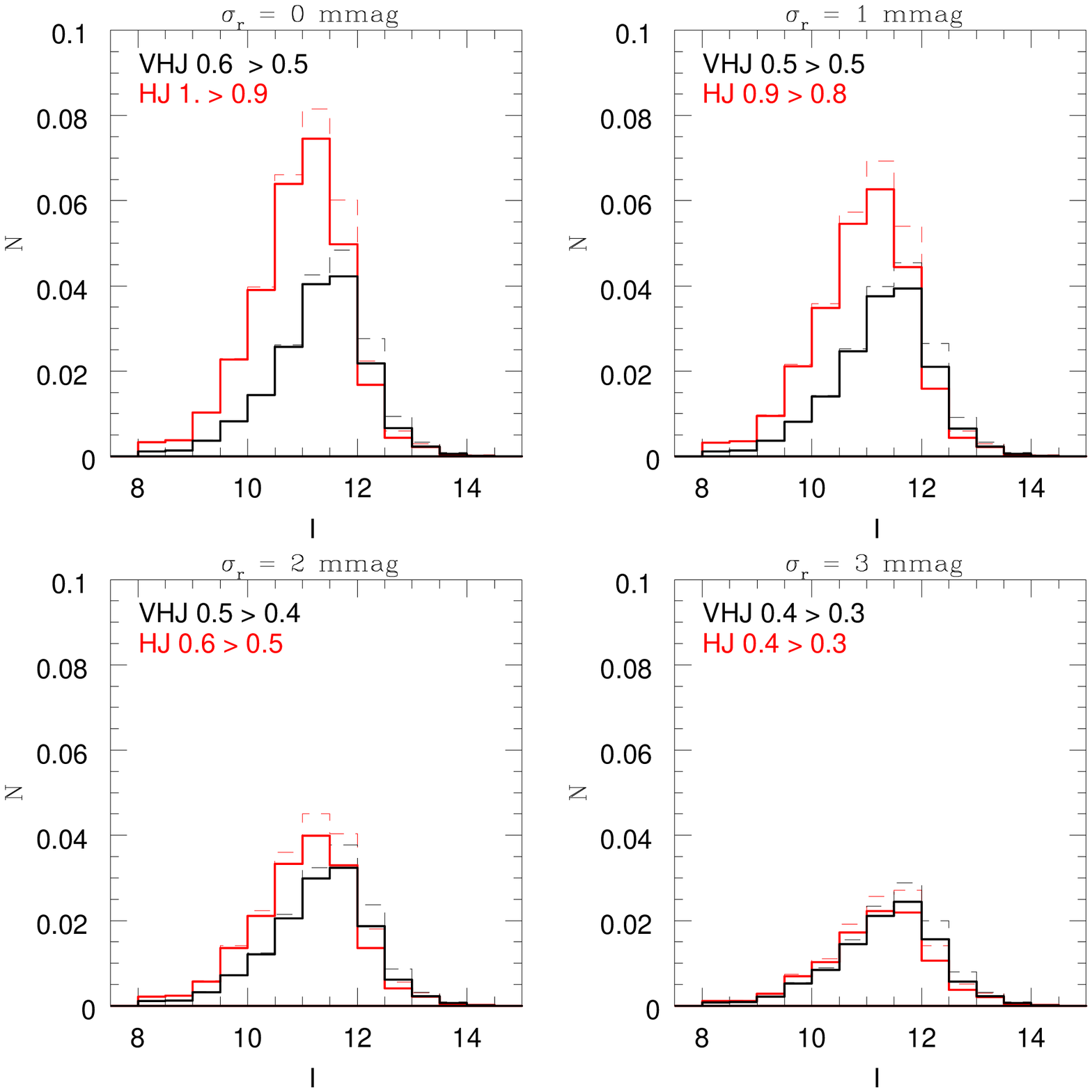}
\caption{Same as Fig. 1, but assuming a continuous observational window as could be obtained from Dome C.}
 \end{figure}

The harvest of our fictitious survey can be judged as very poor, but the existing wide shallow surveys use many telescopes at the same time to cover a large fraction of the sky. Some are also operating from several locations at different longitudes to improve their window function.

\subsection{Deep survey: OGLE-III}

We compute the harvest of the first 2 seasons of the OGLE-III survey. For our computations, we use the actual observation window (A.  Udalski,  private  communication). We assume aperture photometry (in fact, OGLE uses image subtraction + PSF fitting photometry, this point is addressed in G07), the influence of blends has thus to be considered here as somewhat pessimistic. Figures 3 and 4 show our results.

\begin{figure}
\centering
\includegraphics[angle=0,width=9cm]{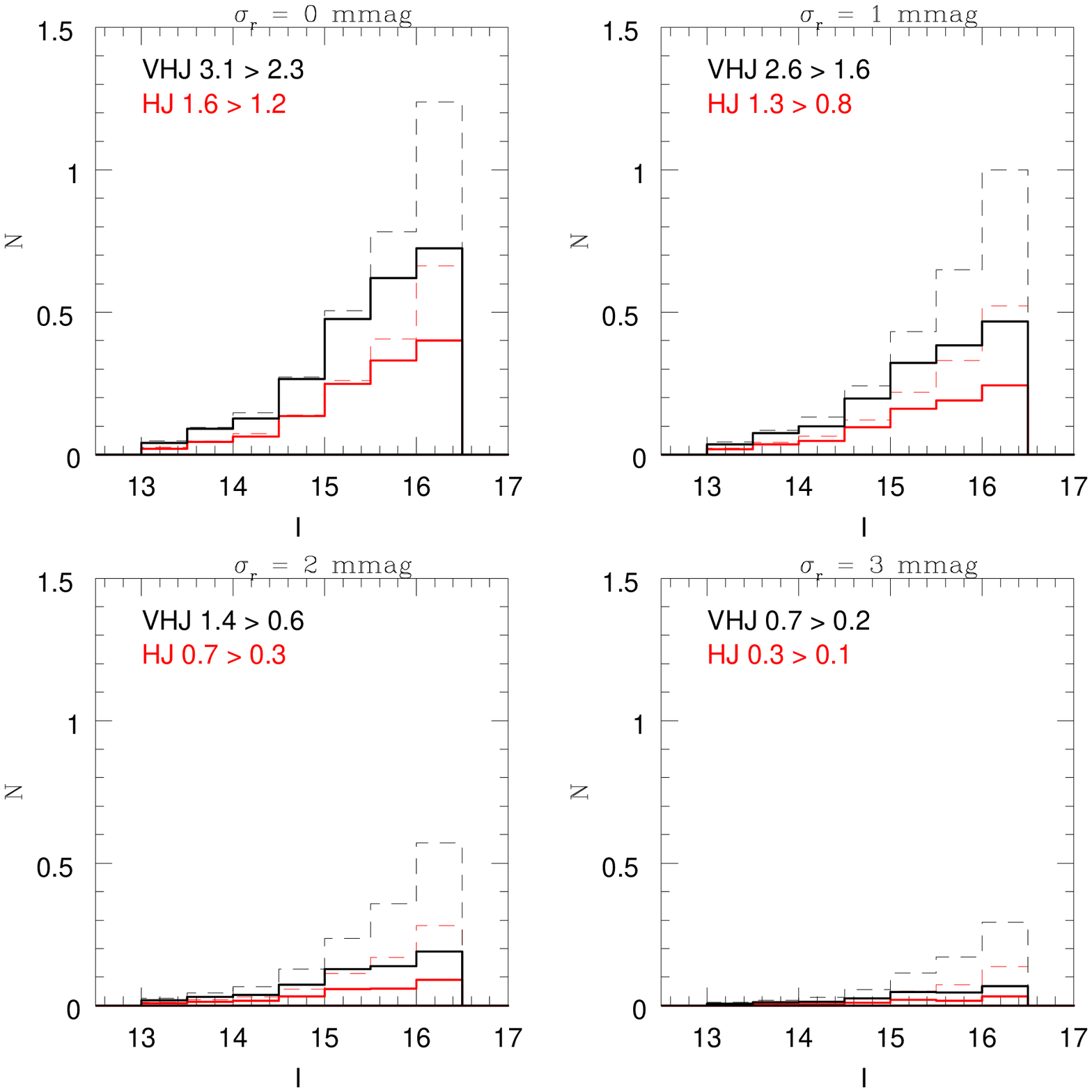}
\caption{Distribution of the Jupiter-like planets detected by OGLE-III-1 (Galactic Bulge) as a function of the $I$ magnitude, for 4 different levels of red noise.}
\end{figure}

\begin{figure}
\centering
\includegraphics[angle=0,width=9cm]{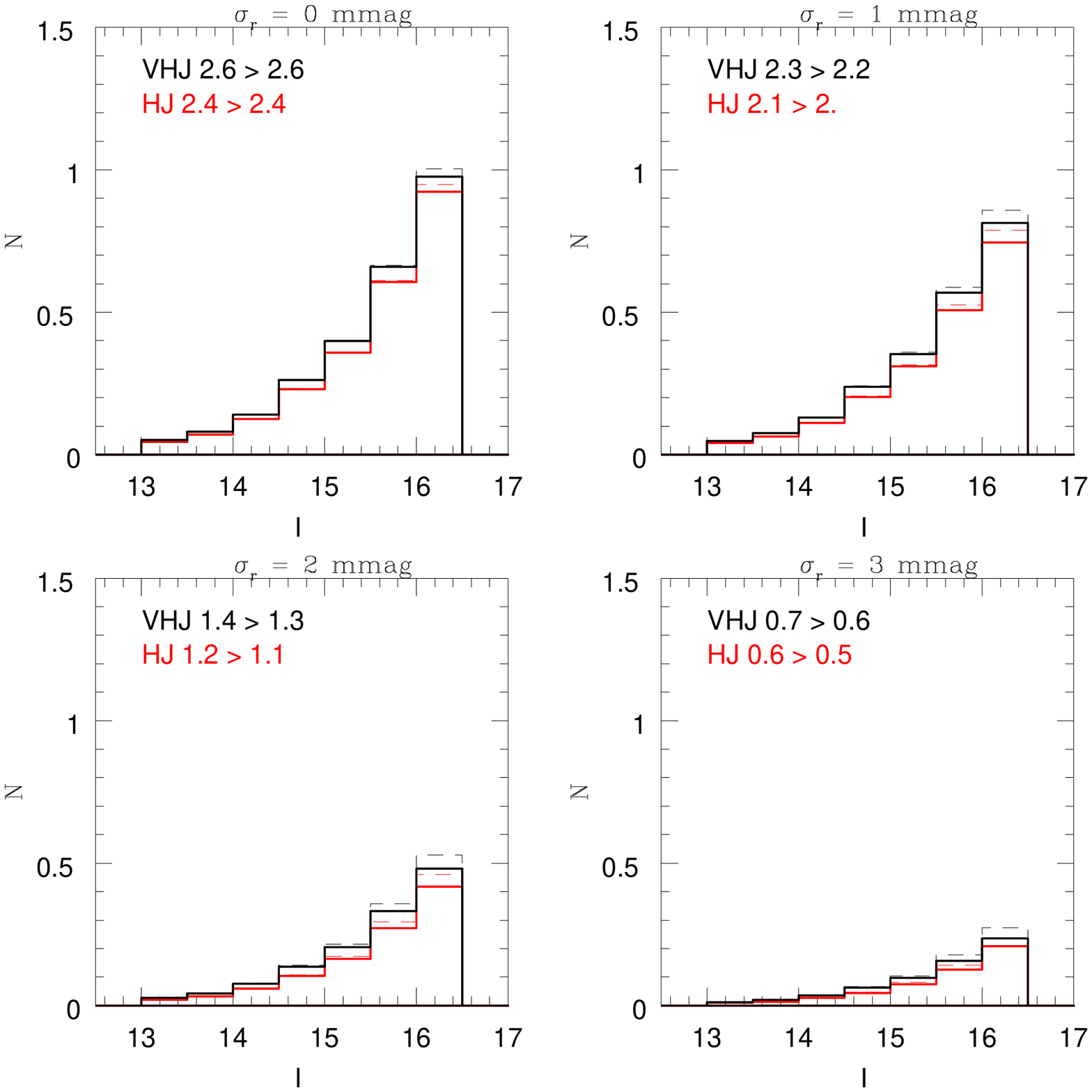}
\caption{Same as Fig. 3, but for OGLE-III-2 (Carina field).}
 \end{figure}

In the case of an extreme crowding (Galactic Bulge fields), the influence of blends appears to be comparable to the one of red noise, at least for moderate red noise, while it is much lower for a disk field such as Carina. The crowding has a larger impact for fainter stars (as would have an increase of the sky background). We also notice that red noise has a moderate influence for amplitudes up to 1 mmag, but that it predominates at larger amplitudes. This outlines the importance of reduction and post-reduction methods used to obtain the light curves.    

\section{Conclusion}

 Our results show that the influence of blends is rather negligible for most transit surveys when compared to the influence of systematics. The only  fields of view leading to a rather large impact of the blends are the highly crowded Bulge fields of the first season of OGLE-III. Nevertheless, one has to keep in mind that part of the red noise \emph{could be due to the crowding}. Indeed, the blends of the PSFs in crowded fields evolve at the same frequency as the seeing, i.e. at a low frequency. If the stellar flux measurements and the quality of the  sky subtraction depend on the level of crowding, as is the case with most reduction methods, it gives rise to the presence of covariant structures within the obtained light curves which would be only partially corrected by detrending algorithms. This aspect of the influence of blends is analyzed in G07. 

\acknowledgements 
We wish  to thank  A. Udalski for providing useful informations about OGLE-III.


\end{document}